# High-resolution computed tomography with scattered x-ray radiation and a single pixel detector

A. Ben Yehuda[1], O. Sefi[1], Y. Klein[1], R. H Shukrun[1,3], H. Schwartz[1], E. Cohen[2], and S. Shwartz[1*]

[1] Physics Department and Institute of Nanotechnology and Advanced Materials, Bar-Ilan University, Ramat Gan, 52900 Israel.

[2] Faculty of Engineering and Institute of Nanotechnology and Advanced Materials, Bar-Ilan University, Ramat Gan, 52900 Israel.

[3] Radiation Safety Department, Soreq Nuclear Research Center, Yavne 81800, Israel.

*Corresponding author. Email: Sharon.shwartz@biu.ac.il

**Abstract:** X-ray imaging is a prevalent technique for non-invasively visualizing the interior of the human body and opaque instruments. In most commercial x-ray modalities, an image is formed by measuring the x-rays that pass through the object of interest. However, despite the potential of scattered radiation to provide additional information about the object, it is often disregarded due to its inherent tendency to cause blurring. Consequently, conventional imaging modalities do not measure or utilize these valuable data. In contrast, we propose and experimentally demonstrate a high-resolution technique for x-ray computed tomography (CT) that measures scattered radiation by exploiting computational ghost imaging (CGI). We show that our method can provide sub-200 μm resolution, exceeding the capabilities of most existing x-ray imaging modalities. Our research reveals a promising technique for incorporating scattered radiation data in CT scans to improve image resolution and minimize radiation exposure for patients. The findings of our study suggest that our technique could represent a significant advancement in the fields of medical and industrial imaging, with the potential to enhance the accuracy and safety of diagnostic imaging procedures.

Commercial x-ray scanners utilize the same physical principle that Wilhelm Roentgen demonstrated in 1895 when he used x-rays to image the hand of his wife (*1*). The x-rays passing through the object are absorbed to varying degrees by the different structures, creating an image that shows the internal composition of the object. In medical imaging, for instance, denser structures such as bones absorb more x-rays and appear white on the image, while softer tissues, like muscles and inner organs, absorb fewer x-rays and appear darker. This very simple concept has proven useful and robust, making x-rays one of the most valuable medical and industrial imaging modalities. However, it became apparent soon after x-rays were first used for imaging that the image quality was significantly degraded in many practical scenarios where the density variations of the organs were small or when the volume of the object was significant (*2–5*). When x-rays interact with the electrons in the object, they are not only absorbed but are also scattered, introducing significant blurring and distortions of the images. To partially mitigate the impact of scattered radiation on the image quality, grids or collimators are used, but these tools also increase the radiation dose leading to a significant increase in the risk of radiation damage (*6–11*). Notwithstanding the successful application of those devices, their ability to reduce scattering is limited and might be insufficient for scenarios where the contrast of the image is low or when a detailed image of the object is required. Despite numerous endeavors to improve image quality and reduce radiation dose, scattering remains a persistent challenge (*12–22*), especially in CT scans since they target large volumes and low contrasts.

Compton scattering in human tissues is stronger than the absorption for the photon energy range typically used for CT scans (80-150 keV), as illustrated in Fig.1 (*23*), which shows the absorption (the photoelectric effect) and the Compton scattering (the incoherent scattering) in cortical bones as a function of the photon energy of the x-ray beam. This plot clearly shows that current x-ray imaging techniques leave substantial amount of energy unused, necessitating higher doses to achieve adequate image quality (*24*, *25*).

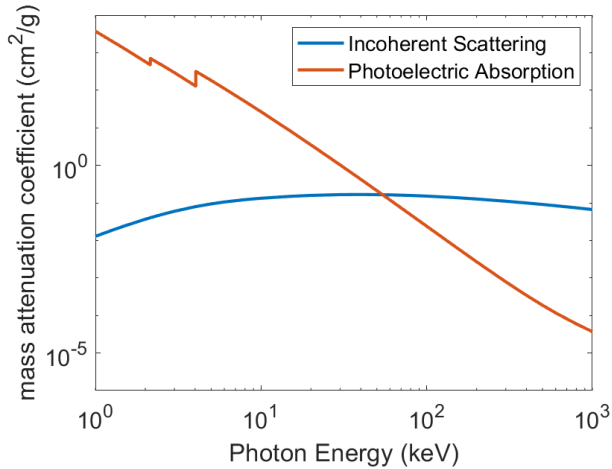

**Fig. 1. Photon energy dependence of the cross-sections of photo absorption (orange) and Compton scattering (blue) in cortical bones.** The incoherent scattering is stronger than the photoelectric absorption for photon energy range of 80-150 keV.

Here we propose and experimentally demonstrate a technique for CT that measures the scattered x-ray radiation by utilizing a structured illumination (SI) approach. We combine computational ghost imaging (CGI) with a recently developed deep learning algorithm (*26*), and an advanced CT image reconstruction toolbox (*27*) to reconstruct a high-resolution three-dimensional (3D) CT image of a part of a large mammal bone. Our results demonstrate sub-200μm image resolution, superior to most commercially available x-ray imaging modalities, and higher image contrast than obtained by direct (transmission-based) imaging.

We chose to employ CGI, which is a form of single-pixel imaging method since it has been shown to be more robust to scattering than standard imaging methods (*28*, *29*). This is because in CGI the spatial information is derived from the spatially modulated beam, which interacts with the object, rather than from the detector pixels. While x-ray CGI has been demonstrated with transmitted (*30–33*), fluorescent (*34*), and refracted radiation (*35*), we apply it to incoherent scattered radiation.

It is worth mentioning that although cameras based on the Compton effect are available (34, 35), their resolution is poor due to the tendency of scattered radiation to blur. As a result, they are not suitable for medical imaging, unlike our proposed method which provides higher resolution and better image quality (*35*, *36*).

In CGI, the input beam is spatially modulated to produce SI on the object. The signal from the object, which can be the intensity transmitted, refracted, or scattered, is then detected by a single-pixel detector and recorded. The measurement is repeated for different SI patterns and the image is reconstructed by solving an inverse problem defined by the equation:

$$Ax = S, \qquad (1)$$

where the mask patterns are represented by the matrix *A*, for which every row is a single illumination structure, the vector *x* is the response function of the object (either the transmission, the reflection, or scattering), and *S* is the detected signal. The reconstruction of the scatter image is done by solving the equation for the vector *x*.

To expedite the reconstruction process and minimize measurement time, we employed a reconstruction algorithm based on the technique of Ghost Imaging using Deep neural network Constraint (GIDC). The algorithm, for the technique was developed to leverage an untrained, self-supervised deep neural network (DNN) to generate far field super-resolution with visible light (*26*). In the present work we were able to modify GIDC to work with x-ray scatter radiation .The approach is based on the concept that a DNN with randomly initialized weights can recover an image more accurately by adding a conventional regularization term, such as the total-variation (TV) (*36*). The minimization of the loss function of GIDC is crucial in order to obtain a high-quality image. The weights of the DNN are adjusted in each iteration with the constraint of the pre-determined TV value. The loss function is reformulated as follows:

$$T_{\varphi*} = argmin_\varphi |AT_\varphi(x_{DGI}) - S|^2 + \tau\mu[T_\varphi(x_{DGI})], \qquad (2)$$

where $T_{\varphi*}$ is the DNN, defined by a set of weights and biases parameters $\varphi$. The objective of GIDC is to find an optimal configuration $\varphi^*$ for the neural network, which effectively constrains the network output to produce a 1D sequence $\tilde{S} = T_\varphi(x_{DGI})$, in accordance with the GI image formation physics as described by Eq. 1. This reconstructed sequence should closely resemble the experimentally acquired bucket signal *S*, where $x_{DGI}$ represents the *x* with normalized illumination patterns. $\mu$ represents TV coefficient and $\tau$ denotes the regularization parameter. This procedure provides a two-dimensional image that contains 16384 pixels constructed using 3468 iterations.

To reconstruct a 3D image using tomographic reconstruction, projections of the object from different angles are essential. To achieve this, we repeated the procedure multiple times while rotating the object, ultimately reconstructing the 3D image from 28 different angles.

We implemented the GI with our scattered radiation scheme using the experimental setup illustrated in Fig. 2. It includes a 80kVp cone beam x-ray source, a 500 mm collimator (not shown) that is mounted between the source and the object to minimize scattering from the surrounding environment, a silver mask that modulates the beam to create structured illumination on the object, a single-pixel scintillation detector for measuring the radiation scattered from the object, and a 360° rotation stage (not shown) for rotating the object during the measurements. The distance between the source and the mask is 1300 mm and the object is located 50 mm downstream of the mask. The spot size of the beam at the mask is about 17 x 19.5 mm².

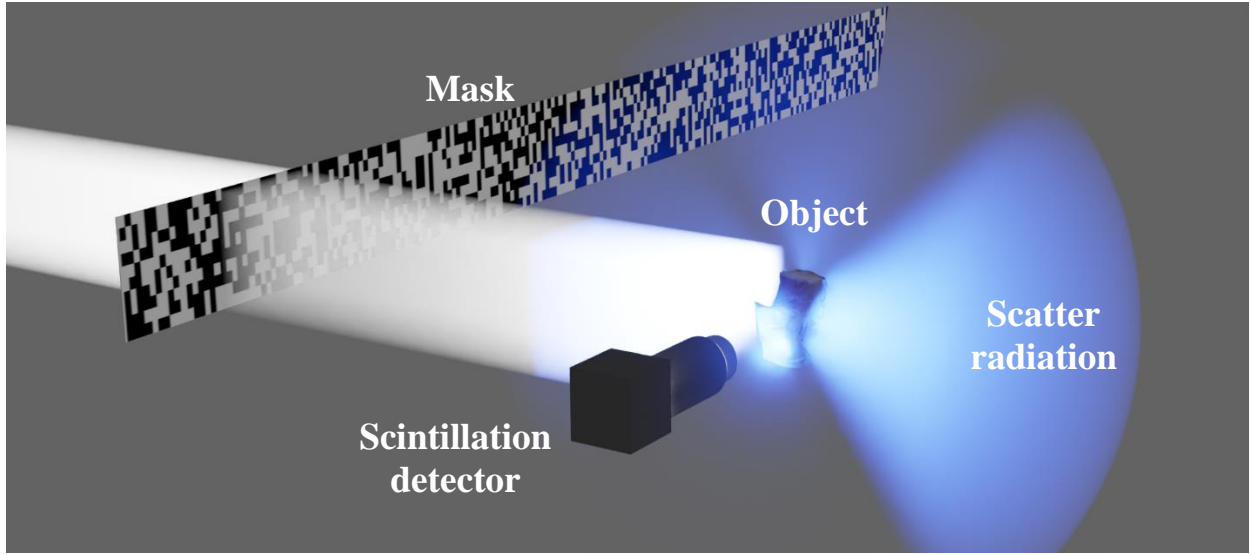

**Fig. 2. Schematic of the experimental setup.** The silver mask modulates the beam, which irradiates the object (bone) with known random binary speckle patterns. The scattered x-ray radiation is then collected by a single pixel scintillation detector and the resulting signals are used to reconstruct a computational image from different angles as the object is rotated using a rotation stage (not shown).

The mask patterns are known randomly distributed binary patterns with transmission-absorption ratio of 1:1, that are irradiated onto the object. High resolution x-ray SI techniques require changing the incident beam for having high contrast and small feature size which can be challenging at high photon energies, as photon penetration depth increases. To overcome this challenge, we adopted a technology made for electrical circuits 3D printing as it can provide high aspect ratio, 100µm printing resolution with silver (see Supplementary Materials).

The scintillation detector is located 50 mm from the object. To demonstrate the ability of imaging from any angle, the detector was mounted at approximately 90° relative to the input beam since the differential cross section given by the Klein-Nishina formula at this angle is small (*37*):

$$\frac{d\sigma}{d\Omega} = \frac{1}{2}r_e^2 \left(\frac{\lambda}{\lambda'}\right)^2 \left[\frac{\lambda}{\lambda'} + \frac{\lambda'}{\lambda} - \sin^2(\theta)\right], \tag{3}$$

where $\frac{d\sigma}{d\Omega}$ is the differential cross section, $\lambda$ is the wavelength of incident x-ray photon, $\lambda$' is the wavelength of scattered x-ray photon, $\theta$ is the scattering angle of the scattered photon and $r_e^2$ is the classical electron radius.

We first calibrated the CT reconstruction algorithm (*38*) by adjusting to the variables which provide the best-quality 3D CT image, with 28 projections we acquired with transmission x-ray. Next, we used these variables to reconstruct the 3D CT image with the corresponding projections we acquired from the scattered x-ray radiation.

To validate our method and to benchmark its efficiency, resolution, and sensitivity against the transmission-based CT reconstruction, we reconstructed 3D images of a bone using signals obtained from transmission, scattering, and their combination. The results of these reconstructions are presented in Figs. 3-5.

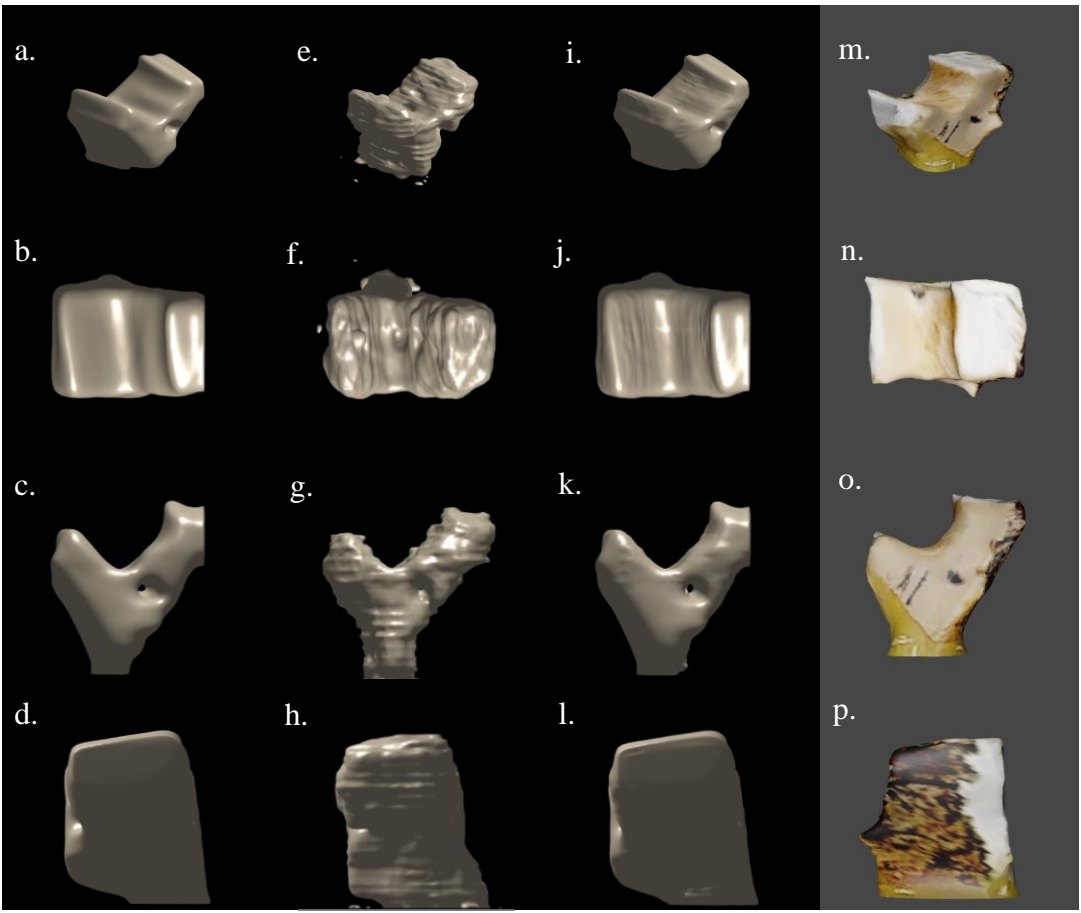

**Fig. 3. Isosurface of the bone from various angles.** The isosurface was reconstructed using data obtained by different imaging modalities: (a)-(d) transmission, (e)-(h) scatter, and (i)-(l) combination of transmission and scattering. The transmission isosurface provides a general representation of the structure of the object while the scatter isosurface offers greater details thanks to its higher resolution and resilience to scattering noise. The combined isosurface shows the general structure of the object maintaining high resolution and displaying the transmission image with additional details. As a guide we added the corresponding 3D LiDAR (laser imaging, detection, and ranging) scans captured with iPhone pro (m)-(p).

In Fig. 3 we present the isosurface of the bone from the reconstructed images using all three methods and compare it to a LiDAR 3D image. It is evident that the scatter reconstruction reveals finer features on the surface of the object compared to the transmission reconstruction, indicating that our method has a higher resolution. The resolution of the images obtained from the transmission data is expected to be approximately 0.5 mm, which is the limit of our detector. In contrast, the resolution of our method is estimated by the autocorrelation width of the mask used to generate the SI and the reconstruction algorithm (*39*, *40*) (see Supplementary Materials). In our case, this length is 155 µm, explaining the superior resolution observed in the scatter reconstruction. The combined reconstruction reveals features which are blurred in the transmission reconstruction due to scattering. This suggests a new approach to eliminate the need for collimators after the object while maintaining high-quality images despite scattering. Adopting this approach has the potential to significantly reduce radiation exposure as collimators absorb a substantial amount of radiation, which represents lost information that could have been collected as demonstrated in our experiment.

It is important to note that unlike previous works on 3D reconstructions using single-pixel detectors with visible light (*41*, *42*), which recovered only the surface gradients to derive the 3D surface of the object, we reconstructed a 3D volume that contains information about both the internal parts and the surface of the object. This enables us to present tomogram slices from top to bottom of the object as is shown in Fig. 4. In particular, we focused on a small hole in the bone, whose size is approximately 2 mm on one end and 1.2 mm on the other and presented its cross sections (Fig. 4j and Fig. 4k). The images and cross sections demonstrate that the edge of the hole nearly disappears in the transmission tomogram due to scattering, while the scatter tomogram remains resilient to this effect and clearly displays the hole.

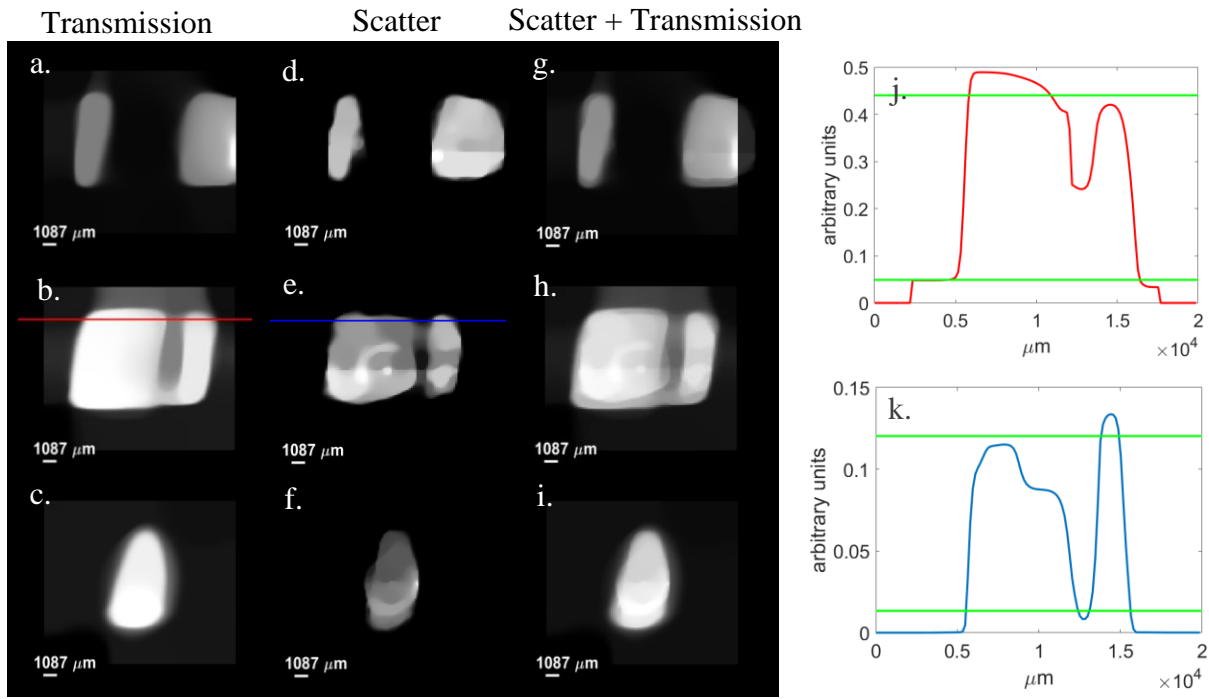

**Fig. 4. Tomogram (cross-sectional) images of the bone sliced top to bottom.** Tomograms from top to bottom: (a)-(c) transmission, (d)-(f) scatter and (g)-(i) a combination of both. The tomograms in each row correspond to the same slice that has been reconstructed using the respective modality. (j) and (k) presents the cross sections of (b) and (e) respectively.

In Figs. 3 and 4 we have demonstrated the potential advantages of our method, but we have also shown several differences between the 3D reconstructions obtained by transition and by scattering. These differences are highlighted in the tomograms of the bone taken from front to back, shown in Fig. 5. One noticeable difference is that with our method the object seems darker in its middle section. This is mostly due to self-absorption and self-scattering that led to a reduction of the signal measured by the detector. Although the differences result from the real physical differences between the mechanisms that govern absorption and scattering, they have to be understood or reconciled to provide the correct 3D structure of the object. This challenge can be mitigated by using another detector (or detectors) at a different angle to compensate for this effect (see Supplementary Materials). We also note that in the tomogram reconstructed by the combination of scattering and transmission the self-attenuation is less pronounced suggesting that the modality can overcome the challenge with some improvement in the algorithms. A second prominent challenge is related to the tradeoff between the measurement time and the image quality. Since our method relies on scanning, the measurement time is proportional to the number of iterations. However, to achieve high image quality, even with clever reconstruction algorithms, the quality of the image is degraded when the number of iterations is much lower than the number of pixels.

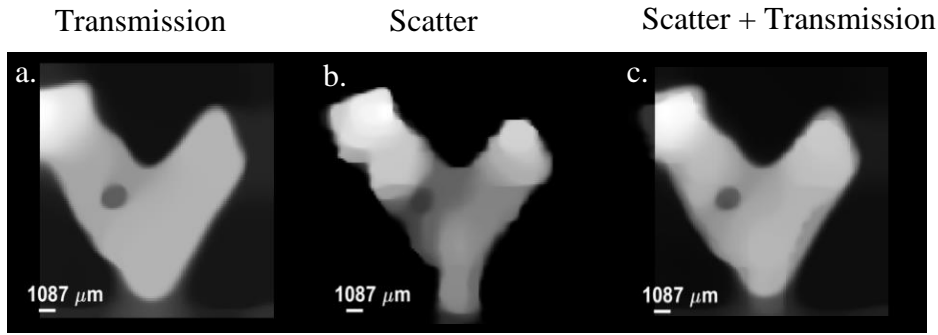

**Fig. 5. Tomogram (cross-sectional) images of the bone sliced front to back.** Tomograms of the front of the bone reconstructed by: (a) the transmission, (b) scatter, and (c) a combination of both.

Our work suggests that scatter radiation can be harnessed to create high resolution CT images, either on its own or in conjunction with traditional transmission information, resulting in a significant improvement in image quality. The selection of approach depends on the specific details of the application and constraints of the measurements. For instance, using scattering-only based CT allows the detector to be mounted at any angle, which can be advantageous in situations where access to certain locations is limited. By combining scatter and transmission information, it is likely that a smaller number of scanning points will be required, making this approach more suitable when dose is a primary concern. Furthermore, our technique has the potential to enhance the image quality of medical CT images by offering higher resolution and lower radiation dose. This is made possible by the simultaneous collection of scattered photons from a 360-degree range. As a result, the need for grids that are currently used in medical x-ray measurements can be eliminated, simplifying the imaging process and further improving image quality. By discarding these grids, we can reduce image artifacts and improve the overall diagnostic accuracy of medical CT imaging.

**Acknowledgments:** Computational support was provided by the NegevHPC project for the simulations.

**Funding:**

This research was supported by the Pazy Research Foundation.

Y.K. gratefully acknowledges the support of the Ministry of Science & Technologies, Israel.

**Author contributions:**

Conceptualization: AB, SS, EC

Methodology: AB, OS, YK

Investigation: AB

Visualization: AB, OS

Monte-Carlo simulations: RHS

Funding acquisition: SS, EC

Project administration: HS

Supervision: SS

Writing – original draft: AB SS

Writing – review & editing: AB, OS, YK, HS, EC, SS

**Competing interests:** Authors declare that they have no competing interests.

# Supplementary Materials for

## High-resolution computational tomography with scattered x-ray radiation and a single pixel detector

A. Ben Yehuda, O. Sefi, Y. Klein, R. H Shukrun, H. Schwartz, E. Cohen, and S. Shwartz[*]

*Corresponding author: Sharon.shwartz@biu.ac.il

**The PDF file includes:**

    Materials and Methods
    Supplementary Text
    Figs. S1 to S5
    Table S1
    References

**Materials and Methods**

Equipment details

The x-ray cone beam source in the experiment was a VJ Technologies P051, which was operated at a voltage of 80kV and a current of 2mA. To collimate the beam, we used a 14 mm diameter, 500 mm long, 18mm thick lead sheet, fashioned into a tube shape, with a slit mounted at its end. The silver mask was custom made using a special 3D printer by Nano Dimension Ltd (43) and it was mounted on two Thorlabs LTS150 stepper motors for precise positioning. We used a Thorlabs K10CR1 stepper motor rotation stage mounted on two Thorlabs MTS50-Z8 stepper motors to rotate the object and align it with respect to the beam. The detector of choice was a Saint Gobain Lanthanum Bromide scintillation detector paired with AS-20 photon multiplier tube. The detector is connected to the Amptek DP5G digital pulse processor and a Canberra 3102D high voltage power supply.

Mask production and calibration

The mask was created using a unique process by Nano-dimension, which utilizes a multi-material multi-layer 3D printer that the company typically uses for fabricating electrical circuits with silver and polymer. We designed the mask to have 1480x1480 pixels with transmission-absorption ratio of 1:1, and dimensions of 160mm x 160mm x 1.5mm with a feature size of 108μm.

To circumvent the potential inhomogeneity in the absorption of the mask, we measured its transmission using a low-resolution flat panel detector. We summed over all pixels in each measurement to normalize the data. To ensure accurate alignment with the manufacturing files on the computer, we used the same low-resolution flat panel detector to capture an image of the mask. We later established a correlation between this image and a low-resolution version of the manufacturing file, which enabled us to accurately locate the mask during the experiment.

Deep neural network reconstruction tool parameters.

Due to the absence of training data of x-ray scattering for high resolution images, we utilized self-supervised deep neural network algorithm for the reconstruction of the GI images. To accomplish this, we utilized the GIDC code, which provides high performance reconstruction but is only able to reconstruct images with number of pixels, that is a power of two in each axis. To meet this requirement, we resized our original image from 184 x 163 pixels to 128 x 128 pixels, resulting in a slight reduction in resolution. Also, GIDC was originally made for Far-field super-resolution reconstructions in visible light. To adapt it to our use we changed the learning rate and the TV strength.

The parameters used in the GIDC code are:

```
img_W = 128
img_H = 128
SR = 0.211                                      # sampling rate
batch_size = 1
lr0 = 0.002                                     # learning rate
TV_strength = 1*10^-3.9                         # regularization parameter of Total Variation
num_patterns = int(np.round(img_W*img_H*SR))    # number of measurement times
Steps = 1501                                    # optimization steps
```

Computational tomography (CT) reconstruction tool parameters
The parameters we used are:

```
Angles=[0, 180, 90, 270, 45, 225, 315, 135, 22.5, 202.5, 67.5, 247.5, 112.5, 292.5, 157.5, 337.5, 355, 175, 348, 168, 353, 173, 5, 185, 12, 192, 300, 120];
niter=500;
angles=Angles*(pi/180)
geo=defaultGeometry('mode','parallel','nDetector',[128 128]);
imgSART_TV=SART_TV(projections,geo,angles,niter,'TVlambda',1000,'TViter',1000);
```

Reconstruction computer specifications:
We used an intel core i7-10700 CPU combined with Nvidia Quadro 2200p GPU and 128GB of RAM.

Simulation details

Simulations were performed using the Monte-Carlo simulation code FLUKA 4-3.1 (44, 45) and the Flair 3.2-2 (46) graphic user interface. We implemented our GI system within the code by modeling each laboratory component separately. The simulation consists of 3 elements: radiation source, object, and detectors (Fig.S1). Technical information about the parameters of each element is described below. A detailed method of implementing GI systems via Monte Carlo simulations can be found in (47).

For this simulation, 2000 realizations were performed, each consisting of 1 million primary particles, to a total of 2 billion particles. The simulation system consisted of 3 components:

- Radiation source – Due to computational considerations, we defined a spatially different radiation source by sampling initial positions of each primary photon from an intensity matrix. Each matrix was set to be a binary matrix with 50*50 pixels, 1.7 cm*1.7cm in size. For each realization a different matrix was used. Then the energy of the primary photon was sampled from energy spectra of an x-ray source. The 80kVp energy spectra was obtained with the SPEKTR 3.0 program (48) using the TASMICS algorithm. We considered a Fewell tube with a tungsten anode and inherent 1.6 mm Al filtration, without added filtration. The source was placed 5 cm from the object.
- Simulation geometry- we defined our object as a 0.85 cm radius sphere mode of bone. The density and elemental composition of the bone were defined as the compact bone material from the Adult Reference Computational Phantoms presented in the ICRP publication 110 (49). The Imaging system was set in the air. Dimensions and material descriptions of each element in the imaging system are presented in table 1.
- Detectors – five detectors were defined, at $0^0$, $45^0$, $90^0$, $135^0$ and $270^0$ with regard to the imaging axis. Each detector was set as a 1.7*1.7 cm ideal boundary crossing detector. We used the USRBDX card in FLUKA that can count the particles crossing a boundary. We defined a one-way fluence detector counting the total of all passing photons.

Spatial resolution

The spatial resolution of our method can be evaluated by measuring the width of the autocorrelation function of the mask that modulates the input x-ray beam (50). The autocorrelation function of the mask is presented in Fig. S2(b), while the 1D horizontal and vertical projections are presented in Figs. S2(c) and S2(d), respectively. The autocorrelation function is nearly isotropic and the FWHM of the curve are 175 µm and 127 µm for the horizontal and vertical axes, respectively.

Experiment at lower photon energy

To validate our method, we also performed an additional experiment with an x-ray tube that emits radiation at 20 keV.
Since we used sandpaper for the mask, we measured it without the object and registered the profiles of the intensity at the object plane by acquiring the images at the various positions of the mask. The experimental setup of this step is described in Fig. S3(a), In the second step, which is depicted in Fig. S3(b), we inserted the object, the figure 7 made from a polylactic acid (PLA) as shown in Fig. S3(c). We measured the x-ray scattering with two energy resolving silicon drift detectors (SDDs) located at two different positions as is shown in Fig. S3(b). We then raster scanned the sandpaper and recorded the radiation, which was scattered from the object. Because the number of pixels in this experiment was only 31 by 40, we used the TVAL3 algorithm (51) for the image reconstruction to avoid the reduction in the resolution that is the results of the pixel resize required by the constraints of the GIDC code.
We reconstructed the image of the object by using the scattered radiation that was collected by the detectors for 3000 realizations is shown in Fig. S4. The number of pixels in the image is 1240 and at each realization we measured about $2 \times 10^6$ scattered photons from the object. Based on edge blurring, we estimate that the spatial resolution of our proof-of-principle experiment was about 150µm. This agrees with the width of the measured autocorrelation function Fig. S4 and indicates that further improvement of the spatial resolution is feasible by a simple reduction of the speckle size (52). As shown in Fig. S4 each detector produced a slightly different image reconstruction, with areas containing more absorbing material appearing darker. For example, in Fig. S4(b) the detector is on the right side of the object, resulting in a darker upper section of the image, as more material was present in that area. However, the bottom of the image was less affected by this effect due to the smaller size of the material, which blocked less radiation. By combining the images produced by each detector, we were able to produce a better, more homogeneous image, as we show in Fig. S4(c). The findings of the study highlight the critical role played by the position of the detector in relation to the object. This positioning can yield additional insights and should be considered when reconstructing images.

Simulation results

To validate the results of the low energy experiment we performed a Monte Carlo (MC) simulation at high energy x-ray (80kVp). MC simulations consider the complex light-matter interactions and are able to solve the radiation transfer problem for the relevant geometry and initial conditions. By using such simulation tools, we provide a deeper insight into the physical mechanism of our experimental observation, an insight that could have been impossible to obtain via traditional simulations that consider only geometrical optics. The simulation results strengthen the findings of our experiment and prove that the positioning of the detector has a major effect on the reconstructed image (Fig. S5).

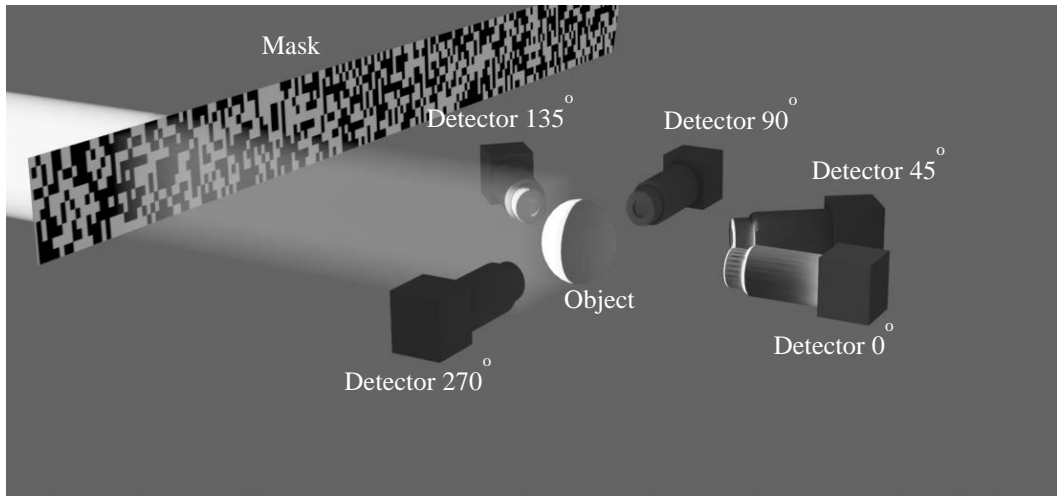

**Fig. S1. Schematic of the simulation setup.** *The simulated mask modulates the beam, which irradiates the simulated object (ball shaped bone) with known random binary speckle patterns. The scattered x-ray radiation is then collected by single pixel detectors mounted at different angles. The resulting signals are then used to reconstruct a computational image from different angles.*

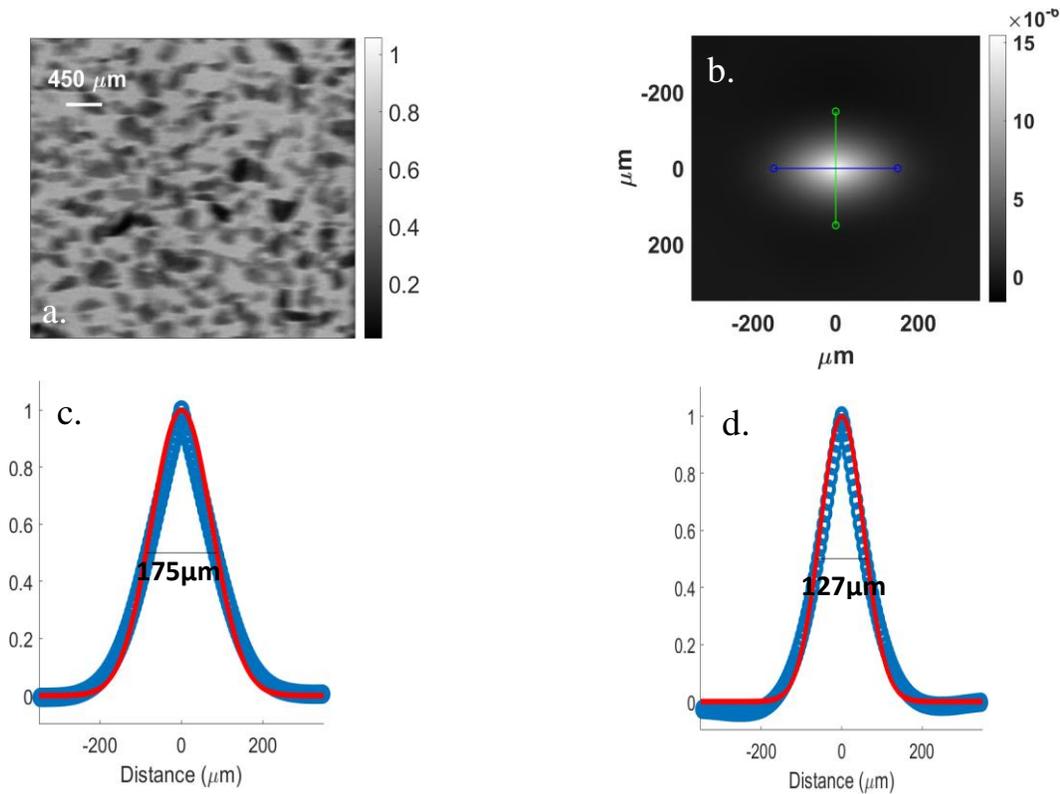

*Fig. S2. Spatial resolution determanation. (a) Example of the reference data (the intensity fluctuations induced by the mask). (b) Autocorrelation function of the intensity pattern induced by the mask. The blue and green lines indicate the horizontal and vertical cross sections shown in (c) and (d). The blue dots are the measured data for the horizontal and vertical cross sections and the red curves are interpolation functions.*

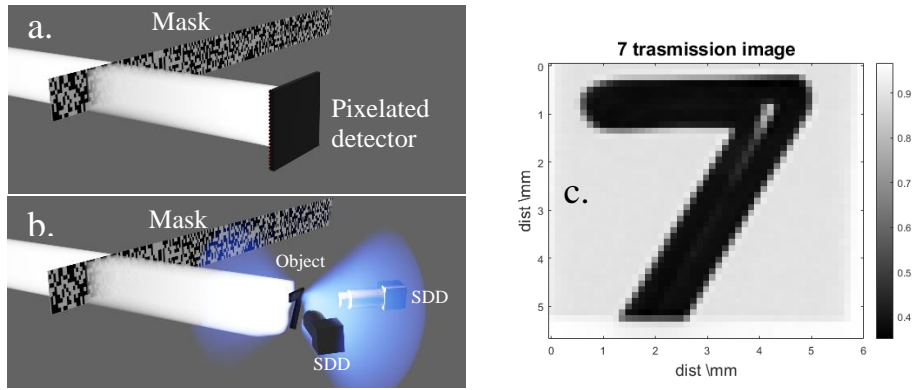

*Fig. S3. Schematic of the experimentl setup and the sample. In step 1 (a), we measure the intensity patterns induced by the mask in the absence of the objects. In step 2 (b) we measure the radiation scattered off the object from both its sides. (c) The image obtained by direct imaging of the object.*

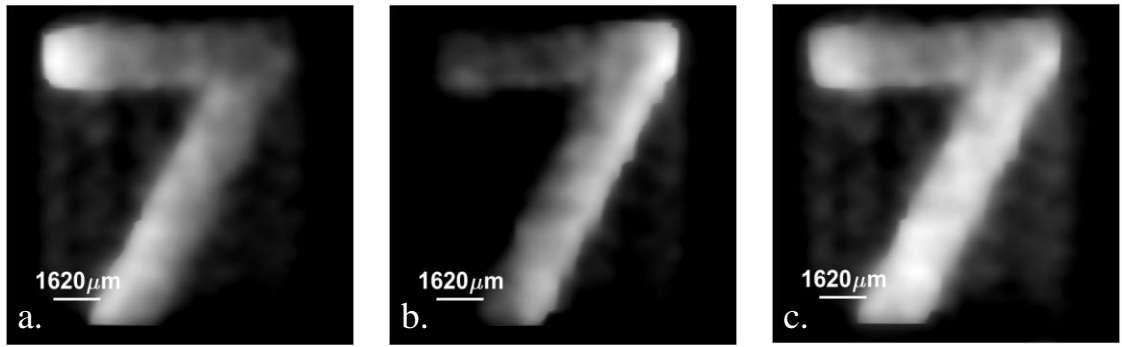

*Fig. S4. Reconstruction images of the object with 3000 ralizations. Image (a) is the reconstruction of the left detector, image (b) is the recontraction of the right detector and image (c) is the combination of both images. This comparison show that different detector position resoult in different image reconstraction due to self absorption and scatter.*

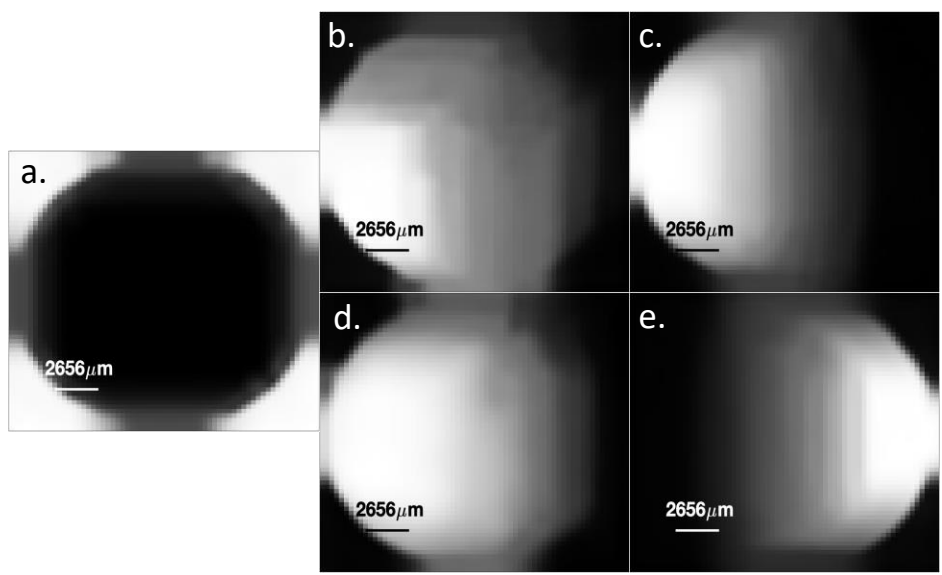

*Fig. S5. Simulation results show the reconstructed images for the detector at: (a) 0 degrees (transmission), (b) 45 degrees, (c) 90 degrees, (d) 135 degrees, and (e) 270 degrees. These simulation results qualitatively agree with our experimental results and support the claim that different detector positions result in different scatter image reconstructions due to physical effects such as self-absorption and self-scattering of the object.*

**Table S1: simulation geometry and materials**

|  | Dimensions (h*w*d) (mm*3) | material | Density (g/cm3) | Elemental composition (% by mass) |
| --- | --- | --- | --- | --- |
| **Object** |  | Bone | 1.85 | H-4.7234, C-14.433, N-4.199, O-44.6096, Mg-0.22, P-10.497, S-0.0315, Ca-20.993 |
| **Surrounding media** | -- | Air | 0.00120484 | C-1.248, N-75.5267, O-23.1781, Ar-1.2827 |